\documentclass[11pt,a4paper]{article}
\usepackage{jcappub}

\usepackage{multirow}
\usepackage[thinlines]{easytable}

\newcommand{\CLASS}{\textsc{class}}

\begin{document}
\hfill CERN-PH-TH-2014-105, LAPTH-046/14
\title{Strongest model-independent bound on the lifetime of Dark Matter}

%
%
%
%

\author[a]{Benjamin Audren,}
\author[a,b,c]{Julien Lesgourgues,}
\author[d]{Gianpiero Mangano,}
\author[c]{Pasquale Dario Serpico,}
\author[a]{and Thomas Tram}

\affiliation[a]{Institut de
Th\'eorie des Ph\'enom\`enes Physiques, \'Ecole Polytechnique
F\'ed\'erale de Lausanne, CH-1015, Lausanne,
Switzerland}
\affiliation[b]{CERN, Theory Division, CH-1211 Geneva 23, Switzerland}
\affiliation[c]{LAPTh, Univ. de Savoie, CNRS, B.P.110, Annecy-le-Vieux F-74941, France}
\affiliation[d]{Istituto Nazionale di Fisica Nucleare - Sezione di Napoli,
Complesso Universitario di Monte S. Angelo, I-80126 Napoli, Italy}

\emailAdd{benjamin.audren@epfl.ch}
\emailAdd{Julien.Lesgourgues@cern.ch}
\emailAdd{mangano@na.infn.it}
\emailAdd{serpico@lapth.cnrs.fr}
\emailAdd{thomas.tram@epfl.ch}

\date{\today}


\abstract{Dark Matter is essential for structure formation in the late Universe so it must be stable on cosmological time scales. But how stable exactly? Only assuming decays into relativistic particles, we report an otherwise model independent bound on the lifetime of Dark Matter using current cosmological data. 
Since these decays affect only the low-$\ell$ multipoles of the CMB, the Dark Matter lifetime is expected to correlate with the tensor-to-scalar ratio $r$ as well as curvature $\Omega_k$. We consider two models, including $r$ and  $r+\Omega_k$ respectively, versus  data from Planck, WMAP, WiggleZ and Baryon Acoustic Oscillations, with or without the BICEP2 data (if interpreted in terms of primordial gravitational waves). This results in a lower bound on the lifetime of CDM given by $160\,\text{Gyr}$ (without BICEP2) or $200\, \text{Gyr}$ (with BICEP2) at 95\% confidence level.}




\maketitle
\section{Introduction}
\subsection{Stability and particle physics}
Although the existence of  {\it dark matter} (DM) is well established by a large number of
observations in cosmology and astrophysics,  we have presently very
few clues on its particle physics nature. This is mostly due to the purely
gravitational origin of the evidence collected so far, which does not provide
any handle for particle identification.
While a number of strategies are ongoing to constrain or detect different
classes of models, it is worth remarking that  already some of the most basic DM
properties can help shedding light on its nature. One such example is provided
by the high stability that this species must possess. If one thinks of
the Standard Model (SM) of particle physics, most of its particles are
unstable: exact stability is in fact the exception and must be enforced by some
exact symmetry, such as the unbroken QED gauge symmetry for the electron or
Lorentz symmetry for the lightest neutrino.  Much more frequent are examples of
meta-stability due to some approximate symmetries, such as for  the heavier
neutrino states, for  the ones often found in nuclear physics (including the
neutron decay) due to mass quasi-degeneracies, and, possibly,  for the proton
itself if the accidental baryon number symmetry is broken at some very high
energy scale as in Grand Unified gauge theories.  In fact, this kind of
considerations provides a useful guideline in DM model building, see
e.g.~\cite{Hambye:2010zb}.

Loosely speaking, one knows that the DM lifetime should be at least
comparable to the lifetime of the universe, otherwise it could not fulfil its
role in structure formation and astrophysical observations. However, inferring
from that phenomenological condition an infinite lifetime is a strong prejudice
dictated by simplicity, but with very little empirical or theoretical
justification. For example, for typical WIMP candidates one often
{\it assumes} a discrete $Z_2$ symmetry under which the SM particles and DM have opposite
charge, but it is easily conceivable that this symmetry  is broken at a more fundamental
level, with the only requirement that the lifetime of the DM particle is
sufficiently long. Stringent bounds on the lifetime $\tau$ of WIMP DM candidates with electroweak scale
masses come, for example, from the diffuse gamma ray flux , at the
level of  $\tau \gg 10^{26}\,$s, see for instance~\cite{Cirelli:2012ut}.
Hence, allowed timescales for the decay should be longer than a billion times
the lifetime of the universe, which would exclude any plausible effect on
gravitational structures.
\subsection{Gravitational effects}
The drawback of these considerations is their model-dependence. In particular,
the bounds depend on the nature and energy distribution of the by-products of
the decay. Interestingly, however, looser but way more general and robust
constraints can be obtained again from purely gravitational considerations.
The key property that allows one to constrain the DM lifetime
gravitationally is that in the decay process, a non-relativistic (usually cold)
DM component is replaced by a combination of radiation and of massive
particles, which in turn have a finite velocity dispersion.  This alters
notably the growth of structures.
More specifically, if significant DM decay takes place, the background evolution of the universe
can show departure from the standard case
and affect several cosmological observables (e.g. the
size of the sound horizon at recombination).  At the perturbation level, one also expects an
enhancement of the Late Integrated Sachs-Wolfe (LISW) effect, beyond the one due to the
cosmological constant, as shown in \cite{Ichiki:2004vi} and described in section~\ref{sec:comparison} below.

%
\subsection{Previous works}
In the past decade, several studies have derived constraints on the DM lifetime using cosmological data, starting from
the study of decaying hot neutrino DM in \cite{Adams:1998nr}. The case of decaying cold DM was first analysed by
Ichiki et al.~\cite{Ichiki:2004vi}, who found a 95\% C.L. bound of 52 Gyr using WMAP-1yr
temperature $C_\ell$ data, and assuming decay into fully relativistic species.
Ref.~\cite{Kaplinghat:1999xy} developed the formalism to describe the
cosmological effects of an unstable relic and its relativistic decay products, both at the background and perturbation levels.  Since then, the
bounds have been refined in two ways. First, more data sets on CMB temperature/polarization and on large scale structures have been
included in the analysis. For example, by including WMAP-5yr,  Type Ia supernova data,
Lyman$-\alpha$ forest, large scale structure and weak lensing observations,
Ref.~\cite{DeLopeAmigo:2009dc} obtained a bound of 100 Gyr (and also updated or corrected previous bounds
from~\cite{Lattanzi:2007ux,Lattanzi:2008ds,Gong:2008gi}).  Second,
some more general bounds have been obtained by allowing the daughter particles
to be massive and thus non-relativistic or only mildly relativistic, see for
instance~\cite{Peter:2010au,Peter:2010sz,Huo:2011nz,Aoyama:2011ba,Wang:2013rha}.
Most recently, a detailed formulation of the problem, both in presence of
massless or massive decay products, has been given in~\cite{Aoyama:2014tga}. In
this reference, it has been additionally shown that the impact of $\sigma_8$
constraints are also  important, and that a possible tension between the value of $\sigma_8$ 
inferred from Planck SZ cluster data and the one extrapolated from CMB temperature data 
could be resolved by assuming $\tau \sim 200\,$Gyr and relativistic daughter particles. 
However, this estimate did not account for parameter degeneracies, and relied on the assumption
that Planck SZ cluster results are not affected by systematic errors.
\subsection{Scope and outline of this paper}
In this paper, we aim at updating cosmological bounds on the DM
lifetime with a proper statistical analysis, accounting for degeneracies and
correlations with other cosmological parameters, as well as estimating the {\it
cosmological model dependence} of the bound thus obtained. In particular,
we will check for degeneracies between decaying DM and spatial curvature, since both can have somewhat similar effects on the CMB. We also consider the impact of including or not BICEP2 results~\cite{Ade:2014xna} on B-mode polarisation interpreted in terms of $r$.

In the following, we limit ourselves to the case of relativistic decay products,
leaving the case of non-relativistic species for future investigation. Note
that this case is nonetheless representative of several DM candidates, for which the decay products are either massless, or at least well inside the relativistic regime. This is usually the case, provided that the produced particles have a much smaller mass than the decaying DM matter particle, and that the decay happens reasonably late. The decay products could consist either in non-standard particles, or in standard model neutrinos produced with typical momenta much larger than their mass. A notable case of such a DM candidate is
represented by the majoron $J$, with mass in the keV
range~\cite{Berezinsky:1993fm,Lattanzi:2007ux,Lattanzi:2008ds,Bazzocchi:2008fh,Lattanzi:2013uza}.
In the simplest {\it see-saw}-like models,  the leading decay channel is in two relativistic 
neutrinos. The majoron lifetime is then inversely proportional to the square of
standard active neutrino masses $m_\nu$,
\begin{equation}
\tau_J =  \frac{16 \pi}{m_J} \frac{v^2}{m_\nu^2} ~.
\label{majorondecayrate}
\end{equation}
Here $m_J$ is the Majoron mass, and $v$ the lepton number breaking scale
\cite{Schechter:1980gr}. Bounds on $\tau_J$ can be used to constrain the value
of $v$ as function of the standard neutrino mass scale. Note that while the results of our study apply also to heavier DM candidates producing energetic neutrinos, these scenarios are better constrained using e.g. limits on the neutrino flux in the Milky Way, leading to stronger bounds (exceeding 10$^6$ Gyr, see for instance~\cite{PalomaresRuiz:2007ry}) than what is found by using cosmological data only. On the other hand, the constraints discussed here are basically the only limits applying to dark matter decaying into unspecified, non-standard forms of {\it dark radiation}.

This paper is structured as follows. In Section~\ref{sec:equations} we recall the formalism
describing a cosmological scenario with a decaying DM candidate, both for the
background and perturbation evolution. Note that we present perturbation equations both in the synchronous gauge (the only case treated 
in the previous literature) and in Newtonian gauge, which allowed us to double-check the numerical results we obtained.
We then describe their implementation in the public numerical code
\CLASS{}\footnote{\url{www.class-code.net}}~\cite{Lesgourgues:2011re,Blas:2011rf}. Section~\ref{sec:comparison} contains a short
description of data sets used in the analysis and our results, and in Section~\ref{sec:conclusions} we conclude and give our outlooks.
\section{Equations and implementation}\label{sec:equations}
\subsection{Background equations}
The background density of the decaying cold DM (dcdm) and of the produced decay radiation (dr) is governed by the two equations
\begin{align}
{\rho_\text{dcdm}}' &= -3 \frac{a'}{a} \rho_\text{dcdm} - a \, \Gamma_\text{dcdm} \, \rho_\text{dcdm}~, \label{eq:dcdm}\\
{\rho_\text{dr}}' &= -4 \frac{a'}{a} \rho_\text{dr} + a \, \Gamma_\text{dcdm} \, \rho_\text{dcdm}~, \label{eq:dr}
\end{align}
where $\Gamma_\text{dcdm}$ is the decay rate defined with respect to proper time, and primes denote derivatives with respect to conformal time. In the language of \CLASS{}, $\rho_\text{dr}$ and $\rho_\text{dcdm}$ fall into the category of \texttt{\{B\}}-variables since they must be evolved alongside the scale factor\footnote{\url{www.cern.ch/lesgourg/class-tour/lecture1.pdf}}. Choosing the fractional energy density in decaying DM plus decay radiation today, $\Omega_\text{dcdm}+\Omega_\text{dr}$, \CLASS{} then finds the corresponding initial condition by using a shooting method. However, since the initial scale factor is set dynamically by the code, we must formulate our initial condition such that it is independent of $a$ in the infinite past. Hence, the target of the shooting method is to find the correct value of the DM energy in a typical comoving volume, $E_\text{ini} \equiv a_\text{ini}^3 \, \rho_\text{dcdm}(a_\text{ini})$. At the same time, we fix the initial condition for the density of decay radiation using the asymptotic solution of Eqs.~(\ref{eq:dcdm}, \ref{eq:dr}) for $a$ going to zero.
\subsection{Perturbation equations in synchronous gauge}
At the level of scalar perturbations, the transfer of energy between the dcdm and dr species is encoded into the continuity and Euler equations of the type
\begin{align}
{T_\text{dcdm}^{\mu 0}}_{;\mu} = -C~, & \qquad {T_\text{dr}^{\mu 0}}_{;\mu} = C~, \label{eq:t0} \\
\partial_i {T_\text{dcdm}^{\mu i}}_{;\mu} = -D~, & \qquad \partial_i {T_\text{dr}^{\mu i}}_{;\mu} = D~. \label{eq:ti} 
\end{align}
The coupling terms $C, D$ accounting for the decay of non-relativistic particles take a trivial form in the synchronous gauge comoving with the decaying species dcdm, i.e. in the gauge such that the metric perturbations $\delta g_{00}$, $\delta g_{i0}$ and the velocity divergence $\theta_\text{dcdm}$ vanish. In this gauge, denoted by the index $(s)$,  $C^{(s)}$ is given by the product of the conformal decay rate, the dcdm particle rest mass  and the local value of the number density of these particles. Expanding this quantity in background and perturbations, one gets
\begin{equation}
C^{(s)} = a \, \Gamma_\text{dcdm} \, \rho_\text{dcdm} \left(1 + \delta_\text{dcdm} \right)~.
\end{equation}
In the same gauge, the decays do not create any additional flux divergence, and $D^{(s)}=0$. Note that assuming similar expressions for $C$ and $D$ in other gauges would lead to wrong results. The Euler equation derived from (\ref{eq:ti}) for dcdm in the synchronous gauge (s) then reads
\begin{equation}
{\theta_\text{dcdm}^{(s)}}' = -\frac{a'}{a} \, \theta_\text{dcdm}^{(s)} = 0~.
\end{equation}
Given adiabatic initial conditions there is no reason for ordinary DM (cdm) and dcdm not to be aligned at early times. Hence, one can fully specify the synchronous gauge by choosing an initial equal-time hypersurface such that $\theta_\text{dcdm} = \theta_\text{cdm}=0$. It follows that they will remain zero at any time and we conclude that the synchronous gauge comoving with cold DM is simultaneously comoving with dcdm. Therefore, one can refer to a single synchronous gauge $(s)$, in which the Euler equations for both cdm and dcdm can be omitted.
\subsection{Perturbation equations in Newtonian gauge}
Since the \CLASS{} code is written in both synchronous and Newtonian gauge, we wish to derive the full set of equations in both gauges, while the previous literature only  presented synchronous equations. Implementing both gauges allows for a useful consistency check, since one must recover the same observables in the two gauges.
After writing the continuity and Euler equations in the synchronous gauge, we gauge-transform them using Eqs. (27a-27b) of \cite{Ma:1995ey}, which take a slightly more complicated form in presence of a decay rate:
\begin{align}
\delta_\text{dcdm}^{(s)} & = \delta_\text{dcdm}^{(n)} + \left(3 \frac{a'}{a} + a \, \Gamma_\text{dcdm}\right) \alpha ~,\\
\delta_\text{dr}^{(s)} & = \delta_\text{dr}^{(n)} + \left(4 \frac{a'}{a} - a \, \Gamma_\text{dcdm} \frac{\rho_\text{dcdm}}{\rho_\text{dr}} \right) \alpha ~,\\
\theta_\text{dr}^{(s)} & =  \theta_\text{dr}^{(n)} - k^2 \alpha = 0~,
\end{align}
with $k$ the wavenumber, $\alpha \equiv (h'+6\eta')/2k^2$ and where we address the reader to~\cite{Ma:1995ey} for the (by now standard) notation of the different potentials.
\begin{table}%
\begin{centering}
\begin{tabular}{l| c c}
                  &  Synchronous                         &  Newtonian \\
\hline && \\
$\mathfrak{m}_\text{cont}$  &  $\dot{h}/2$                 &  $-3\dot{\phi}$  \\
$\mathfrak{m}_\psi$  &  $ 0 $       &  $\psi$ \\ 
$\mathfrak{m}_\text{shear}$  &  $(\dot{h}+6\dot{\eta})/2$  &  $0$
\end{tabular}
\caption{Metric source terms for scalar perturbations in synchronous and Newtonian gauge.}
\label{tab:metric}
\end{centering}
\end{table}
The final set of equations in {\it both} gauges can be written as
\begin{align}
{\delta_\text{dcdm}}' & = - \theta_\text{dcdm} - \mathfrak{m}_\text{cont} - a \, \Gamma_\text{dcdm} \mathfrak{m}_\psi ~, \\
{\theta_\text{dcdm}}' & = -\frac{a'}{a} \, \theta_\text{dcdm} + k^2 \mathfrak{m}_\psi~, \\
{\delta_\text{dr}}' & = - \frac{4}{3} \left(\theta_\text{dcdm} + \mathfrak{m}_\text{cont}\right)  + a \, \Gamma_\text{dcdm} \frac{\rho_\text{dcdm}}{\rho_\text{dr}} \left(\delta_\text{dcdm} - \delta_\text{dr} + \mathfrak{m}_\psi \right) ~, \\
{\theta_\text{dr}}' & = \frac{k^2}{4} \delta_\text{dr} - k^2 \sigma_\text{dr} + k^2 \mathfrak{m}_\psi  - a \, \Gamma_\text{dcdm} \frac{3 \rho_\text{dcdm}}{4 \rho_\text{dr}} 
\left( \frac{4}{3} \theta_\text{dr} - \theta_\text{dcdm} \right)~,
\end{align}
where the metric source terms $\mathfrak{m}_\text{cont}$ and $\mathfrak{m}_\psi$ are given in Table~\ref{tab:metric}.
\subsection{Boltzmann hierarchy for decay radiation}
The full perturbations of the decay radiation distribution function can be written in different ways. We adopt here the same set of equations as in~\cite{Kaplinghat:1999xy},  in which the perturbations of the (integrated) phase-space distribution function are defined as
\begin{equation}
F_\text{dr} \equiv \frac{\int dq q^3 f_\text{dr}^0 \Psi_\text{dr}}{\int dq q^3 f_\text{dr}^0} r_\text{dr}~, \label{eq:Fdr}
\end{equation}
with $r_\text{dr}$ defined as
\begin{equation}
r_\text{dr} \equiv \frac{\rho_\text{dr}a^4}{\rho_{\text{cr},0}}~,
\end{equation}
where the the critical energy density today, $\rho_{\text{cr},0}$, has been introduced to make $r_\text{dr}$ dimensionless. The derivative of $r_\text{dr}$ is given by 
\begin{equation}{r_\text{dr}}' = a \, \Gamma_\text{dcdm} \rho_\text{dcdm} / \rho_\text{dr}~,\end{equation}
so that $r_\text{dr}$ is constant in absence of a source. The point of introducing such a factor in the definition of $F_\text{dr}$ is to cancel the time-dependence $F_\text{dr}$ due to the background distribution function $f_\text{dr}^0$ in the denominator of equation~\eqref{eq:Fdr}. This simplifies the Boltzmann hierarchy for the Legendre multipoles $F_{\text{dr},\ell}$, which obey the following equations
\begin{align}
F_{\text{dr},0}' &= -k F_{\text{dr},1} - \frac{4}{3} r_\text{dr} \mathfrak{m}_\text{cont} + {r_\text{dr}}' \left(\delta_\text{dcdm}  + \mathfrak{m}_\psi\right)~, \\
F_{\text{dr},1}' &= \frac{k}{3} F_{\text{dr},0} - \frac{2k}{3} F_{\text{dr},2} + \frac{4k}{3} r_\text{dr} \mathfrak{m}_\psi + \frac{{r_\text{dr}}'}{k} \theta_\text{dcdm}~, \\
F_{\text{dr},2}' &= \frac{2k}{5} F_{\text{dr},1} - \frac{3k}{5} F_{\text{dr},3} + \frac{8}{15} r_\text{dr} \mathfrak{m}_\text{shear}~, \\
F_{\text{dr},\ell}' &= \frac{k}{2\ell+1} \left( \ell F_{\text{dr},\ell-1} - (\ell+1) F_{\text{dr},\ell+1} \right), \,\,\,\,\, \ell>2~. 
\end{align}
The expression for $\mathfrak{m}_\text{shear}$ can be found in Table~\ref{tab:metric}.
For the sake of simplicity, we have reported these equations in a spatially flat universe, but for our analysis we implemented the equations in a general curved FLRW model, following~\cite{Lesgourgues:2013bra}. The Boltzmann hierarchy is truncated at some $\ell_\text{max}$ following the prescription of~\cite{Ma:1995ey} generalised to spatial curvature~\cite{Lesgourgues:2013bra}.
\section{Comparison with data}\label{sec:comparison}
\subsection{Observable effects} 
When discussing the effect of a given parameter on the CMB describing some new physics, one should specify which other parameters are kept fixed. The best choice is the one allowing to cancel all trivial effects, in order to isolate the distinct residual effect associated to the new physics.

Here the focus is on  the effect of the DM decay rate $\Gamma_\mathrm{dcdm}$. If we were varying $\Gamma_\mathrm{dcdm}$ while keeping the DM density fixed {\it today} (either the physical density $\omega_\mathrm{dcdm}=\Omega_\mathrm{dcdm} h^2$ or fractional density  $\Omega_\mathrm{dcdm}$), the code would automatically adjust initial conditions in the early universe. The direct effect of $\Gamma_\mathrm{dcdm}$ on the perturbations would then be mixed with that of changing the early cosmological evolution, and in particular the redshift of equality.

Hence, a better choice is to fix all initial conditions, so that varying $\Gamma_\mathrm{dcdm}$ only affects the late cosmological evolution. In order to do this easily, we implemented an alternative parametrisation in \CLASS{}. Instead of providing $\omega_\mathrm{dcdm+dr}$ or $\Omega_\mathrm{dcdm+dr}$ as input and letting the code compute the initial dcdm density, the user can choose to pass the initial  density of decaying DM (in dimensionless units, as $\Omega_\mathrm{dcdm}^\mathrm{ini} \equiv (\rho_\mathrm{dcdm}^\mathrm{ini} a^3 / \rho_\mathrm{cr,0})$ or $\omega_\mathrm{dcdm}^\mathrm{ini} \equiv \Omega_\mathrm{dcdm}^\mathrm{ini} h^2$), and the code will find the correct density today. Note however, that this procedure also involves a shooting method in order to satisfy the closure equation $\sum_i \Omega_i = 1 - \Omega_k$. With this approach, we preserve the full cosmological evolution at least until photon decoupling, since for realistic values of $\Gamma_\mathrm{dcdm}$ allowed by observations, the DM decay is only significant at late time, long after photon decoupling. In particular, the effects of $\Gamma_\mathrm{dcdm}$ on the CMB are the following: 
\renewcommand{\theenumi}{\roman{enumi}}%
\begin{enumerate}
\item [i)] a change in the angular diameter distance to decoupling, shifting the whole CMB spectra in multipole space;
\item [ii)] a late Integrated Sachs-Wolfe (ISW) effect, since a modification of the homogeneous and perturbed density of DM  at late times affects the evolution of metric fluctuations through the Poisson equation;
\item [iii)]a different amount of CMB lensing, affecting the contrast between maxima and minima in the lensed CMB spectra.
\end{enumerate}
\begin{figure}[ht]%
\includegraphics[width=\columnwidth]{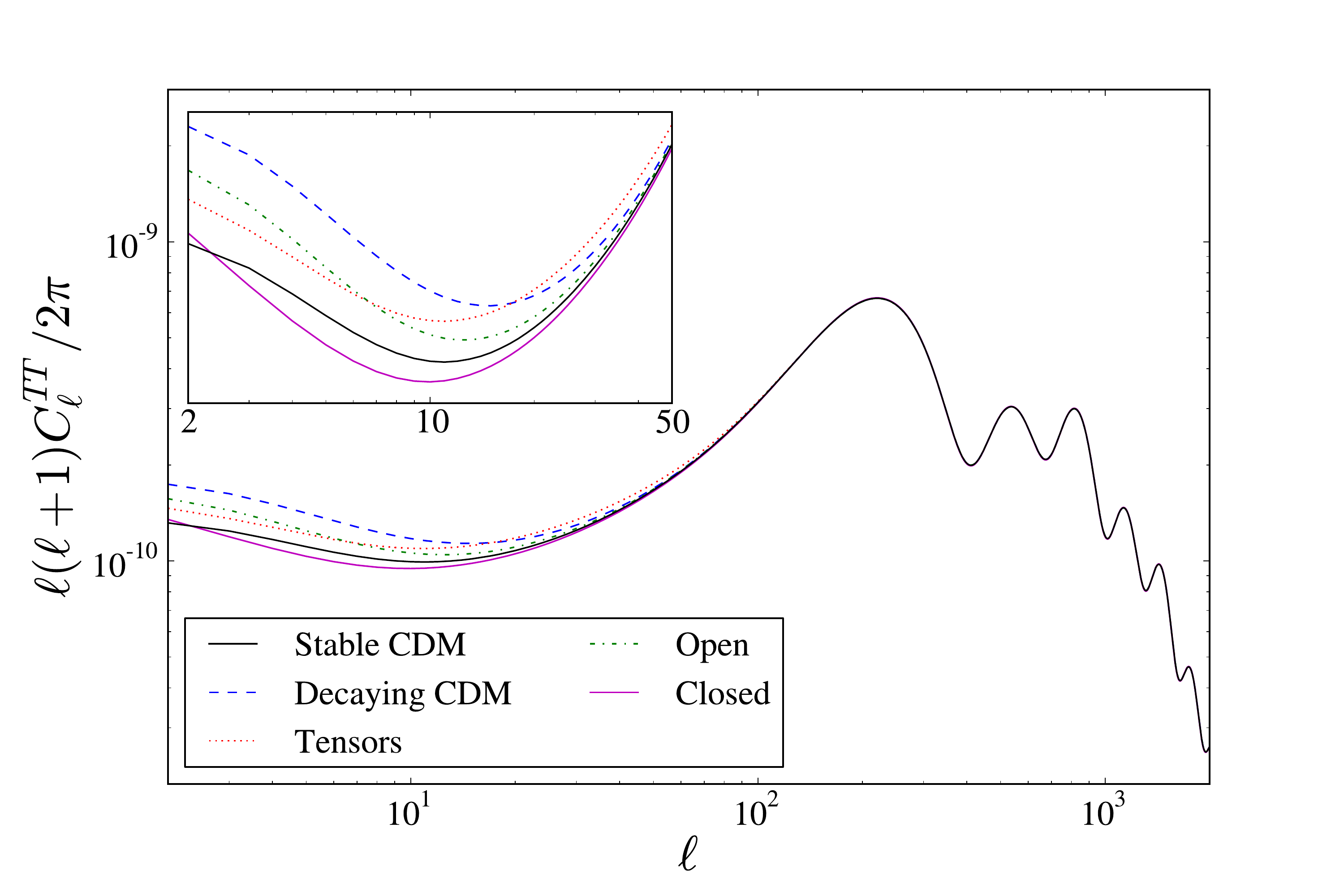}%
\caption{CMB temperature power spectrum for a variety of models, all with the same parameters $\{100\,\theta_s, \omega_\mathrm{dcdm}^\mathrm{ini}, \omega_\mathrm{b}, \ln(10^{10} A_s),  n_s, \tau_\mathrm{reio} \} = \{1.04119, 0.12038, 0.022032, 3.0980,  0.9619, 0.0925\}$ taken from the Planck+WP best fit~\cite{Ade:2013zuv}. For all models except the ``Decaying CDM'' one, the decay rate $\Gamma_\mathrm{dcdm}$ is set to zero, implying that the ``dcdm'' species is equivalent to standard cold DM with a present density $\omega_\mathrm{cdm} = \omega_\mathrm{dcdm}^\mathrm{ini} = 0.12038$. The ``Decaying CDM'' model has $\Gamma_\mathrm{dcdm}=20\,\mbox{km s}^{-1}\mbox{Mpc}^{-1}$, the ``Tensors'' model has $r=0.2$, and the ``Open'' (``Closed'') models have $\Omega_k = 0.02$ ($-0.2$). The main differences occur at low multiples and comes from either different late ISW contributions or non-zero tensor fluctuations.}
\label{fig:cltot}%
\end{figure}
To check (ii), we plot in Figure~\ref{fig:cltot} the unlensed temperature spectrum of models with $\Gamma_\mathrm{dcdm}$ set either to 0 or  $20\,\mbox{km s}^{-1}\mbox{Mpc}^{-1}$~\footnote{It is useful to bear in mind the conversion factor $1\,\mbox{km s}^{-1}\mbox{Mpc}^{-1}= 1.02\times 10^{-3} \mbox{Gyr}^{-1}$. }. To keep the early cosmological evolution fixed, we stick to constant values of the density parameters ($\omega_\mathrm{dcdm}^\mathrm{ini}$, $\omega_\mathrm{b}$), of primordial spectrum parameters ($A_s$, $n_s$) and of the reionization optical depth $\tau_\mathrm{reio}$. Of course, for $\Gamma_\mathrm{dcdm}=0$, the dcdm species is equivalent to standard cold DM with a current density $\omega_\mathrm{cdm} = \omega_\mathrm{dcdm}^\mathrm{ini}$. We need to fix one more background parameter in order to fully specify the late cosmological evolution. Possible choices allowed by \CLASS{} include $h$, or the angular scale of the sound horizon at decoupling, $\theta_s = r_s(t_\mathrm{dec})/d_s(t_\mathrm{dec})$. We choose to stick to a constant value of $\theta_s$, in order to eliminate the effect (i) described above, and observe only (ii). We see indeed in Figure~\ref{fig:cltot} that with such a choice, the spectra of the stable and decaying DM models overlap everywhere except at small multipoles. To check that this is indeed due to a different late ISW effect, we show in Figure~\ref{fig:clsplit} the decomposition of the total spectrum in individual contribution, for the stable model and a dcdm model in which the decay rate was pushed to $100~\mbox{km s}^{-1}\mbox{Mpc}^{-1}$.
\begin{figure}[ht]%
\includegraphics[width=\columnwidth]{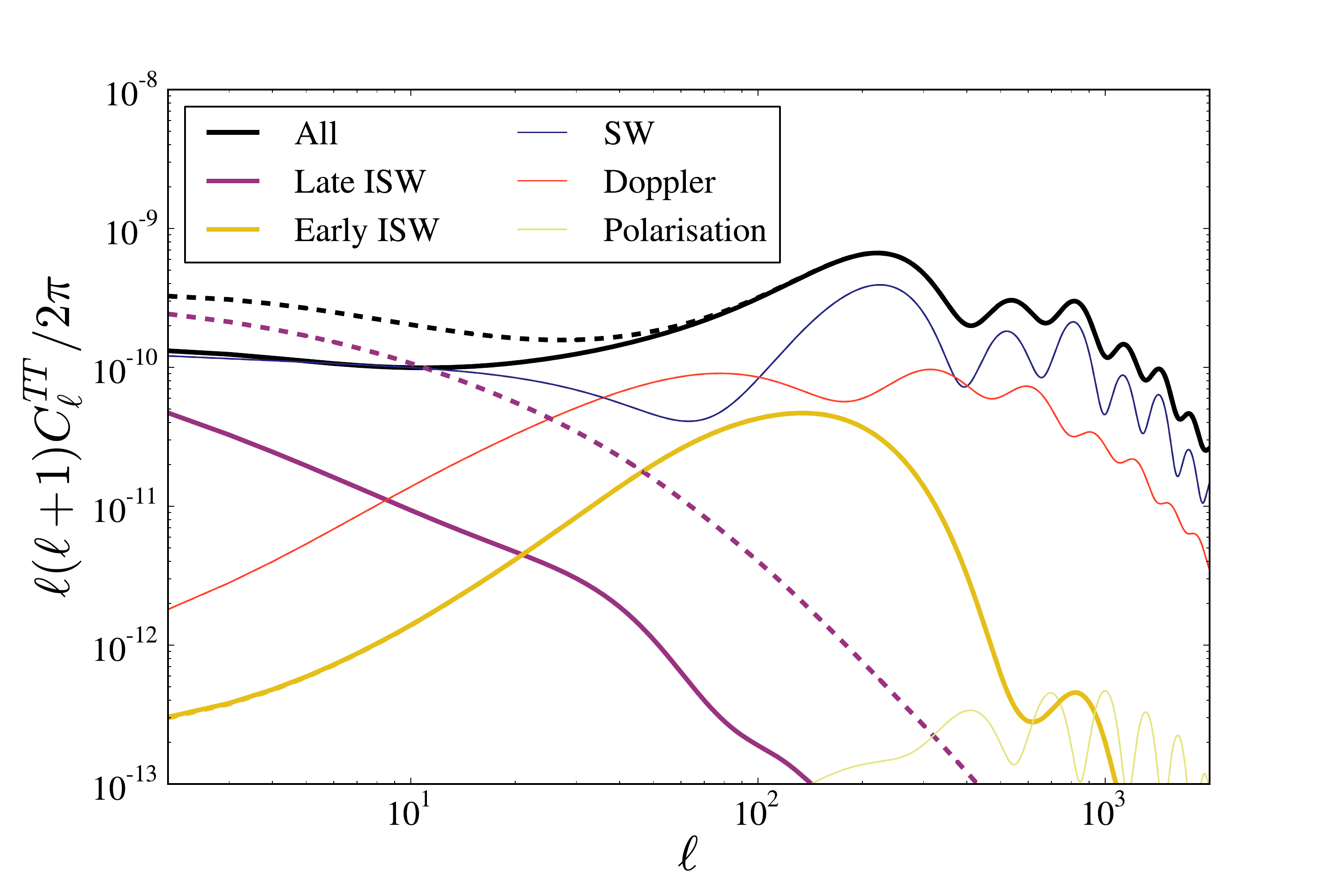}%
\caption{The single contributions to the CMB temperature spectrum (Sachs-Wolfe, early and late Integrated Sachs-Wolfe, Doppler and polarisation-induced) for a stable model (solid) and a dcdm model (dashed) with $\Gamma_\mathrm{dcdm}=100$~km/s/Mpc. The value of other parameters is set as in Figure~\ref{fig:cltot}. We see that only the late ISW effect is sensitive to the decay rate (for other contributions, solid and dashed lines are indistinguishable).}%
\label{fig:clsplit}%
\end{figure}
\begin{figure}[ht]%
\includegraphics[width=\columnwidth]{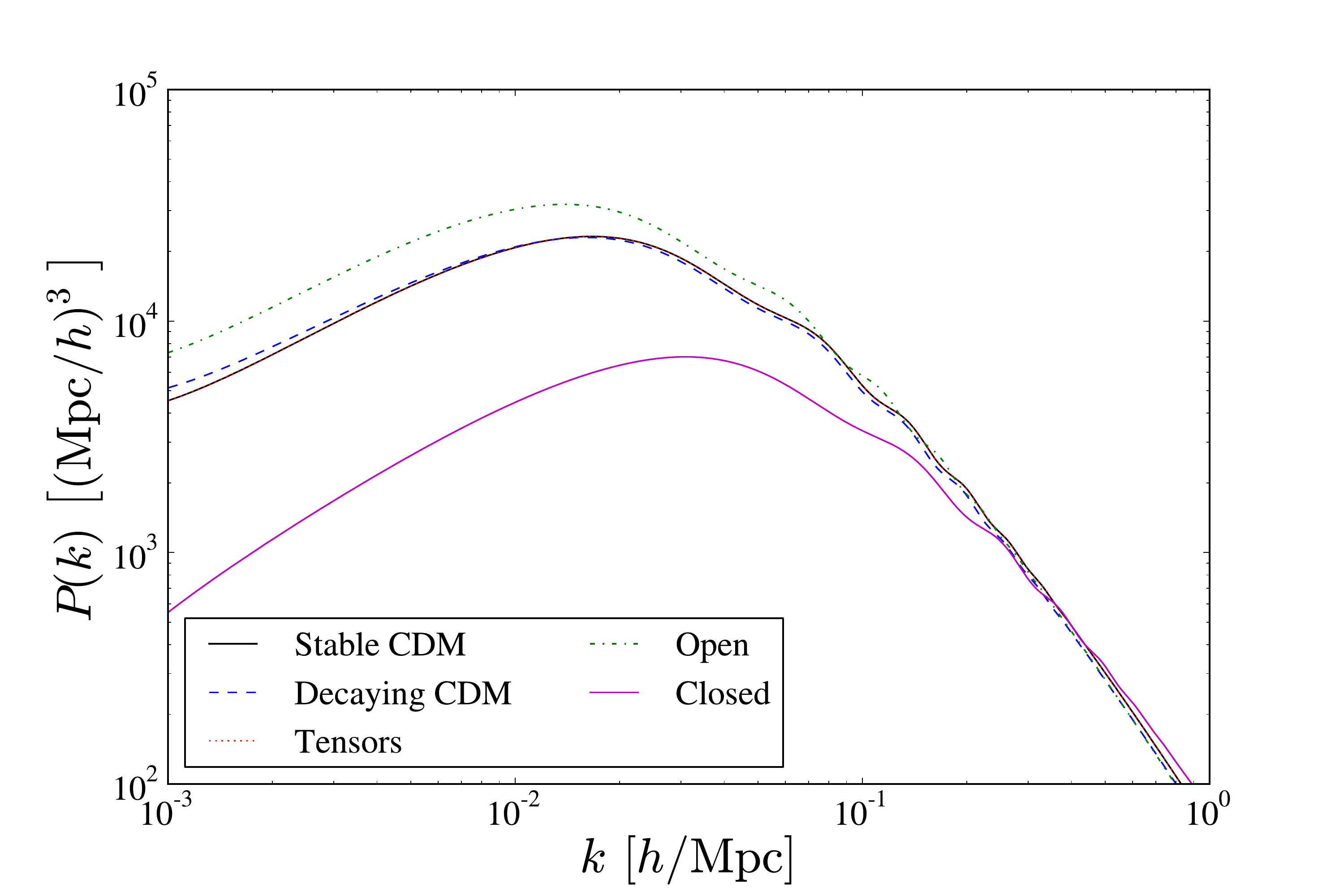}%
\caption{Matter power spectrum $P(k)$ (computed in the Newtonian gauge) for the same models considered in Figure~\ref{fig:cltot}. The black curve (Stable CDM) is hidden behind the red one (Tensors).}%
\label{fig:pk}%
\end{figure}

Since the dominant effect of decaying DM is a modification of the small-$\ell$ part of the CMB temperature spectrum, in the rest of the analysis, it will be relevant to investigate degeneracies between $\Gamma_\text{dcdm}$ and other parameters affecting mainly the large-angle CMB spectra, like the spatial curvature parameter $\Omega_k$ or the tensor-to-scalar ratio $r$ (defined throughout this paper at the pivot scale $k_*=0.05$/Mpc). We show examples of such models in Figure~\ref{fig:cltot}, from which it is not obvious that very small variations of $\Gamma_\mathrm{dcdm}$, $\Omega_k$ and $r$ can be distinguished, given the cosmic variance uncertainty on low $\ell$'s. It is useful to plot the matter power spectrum $P(k)$ of the same models, to see whether CMB lensing or direct measurements of $P(k)$ can help to reduce the degeneracy. This is done in Figure~\ref{fig:pk}. We see that all the parameters discussed here have a different effect on $P(k)$. Playing with tensor modes leaves the matter power spectrum invariant, since it is related to scalar perturbations only. Varying $\Gamma_\mathrm{dcdm}$ changes $P(k)$ slightly for several reasons: 
\begin{itemize}
\item the different background evolution of $\rho_\mathrm{dcdm}$ leads to an overall vertical shift of the spectrum; 
\item the different values of $h$ needed to get the same $\theta_s$ changes the ratio of the Hubble scale at equality and today, hence shifting the spectrum horizontally; 
\item on top of these shifting effects, the different evolution of $\delta_\mathrm{dcdm}$ is such that dcdm has a reduced linear growth factor, affecting the actual shape of the matter power spectrum.
\end{itemize}
When introducing the curvature parameter, one gets a combination of the first two effects only. Moreover, variations of $\Gamma_\text{dcdm}$ and $\Omega_k$ leading to an effect in the CMB of the same amplitude give effects on the $P(k)$ with very different amplitudes. This comparison shows that, at least in principle, CMB lensing effects and direct constraints on $P(k)$ may help to break degeneracies, and to measure $\Gamma_\text{dcdm}$ independently of $\Omega_k$ and $r$. This can only be confirmed by a global fit to current observations.

\subsection{The data}
The parameter extraction is done using a Metropolis Hastings
algorithm, with a Cholesky decomposition to better handle the large number of
nuisance parameters~\cite{Lewis:2013hha}. We investigate two combinations of experiments which we denote by $A$ and $B$. Both share the Planck likelihoods, consisting of the
low-$\ell$, high-$\ell$, lensing reconstruction and low-$\ell$ WMAP polarisation,
as well as the WiggleZ data~\cite{Parkinson:2012vd}, and the
BOSS measurement of the Baryon Acoustic Oscillation
scale at $z=0.57$~\cite{Anderson:2013zyy}. The set $B$ adds the BICEP2 public likelihood code~\cite{Ade:2014xna}. We used the publicly available Monte Python\footnote{\url{https://github.com/baudren/montepython_public}}
code~\cite{Audren:2012wb} for the analysis.

We performed the analysis selecting flat priors for the following
set of parameters
\[\{\omega_b, H_0, A_s, n_s, \tau_{\rm reio},\omega_{\rm dcdm+dr}, \Gamma_{\rm
dcdm}, r, \Omega_k\}~,\]
in addition to the other nuisance parameters for the Planck likelihood, omitted
here for brevity. The first five cosmological parameters stand respectively
for the baryon density, the Hubble parameter, the amplitude at $k_{*}=0.05$/Mpc and tilt
of the initial curvature power spectrum, and the
optical depth to reionisation.  
The next parameter $\omega_{\rm dcdm+dr}$ denote the
physical density of decaying dark matter plus its decay product today (in practise, $\omega_{\rm dcdm+dr}$ is extremely close to $\omega_{\rm dcdm}$ up to typically 4\%). Finally, the last two parameters are the dcdm decay rate and the tensor-to-scalar ratio, also measured at the pivot scale
$k_{*}=0.05$/Mpc. In some of our runs, we vary the curvature parameter $\Omega_k = 1 - \Omega_\mathrm{tot}$.

The tensor tilt $n_t$ is set to satisfy the self-consistency
condition from inflation, {\it i.e} ${n_t=-r/8(2-r/8-n_s)}$, whereas the tensor
running $\alpha_t$ is neglected.  
For the neutrino sector, for simplicity, we performed the same assumption as in~\cite{Ade:2013zuv} (two relativistic neutrinos and one with a mass of $0.06$~eV).
%
\subsection{Results}

\begin{figure}[ht]
\centering
\includegraphics[width=0.6\textwidth]{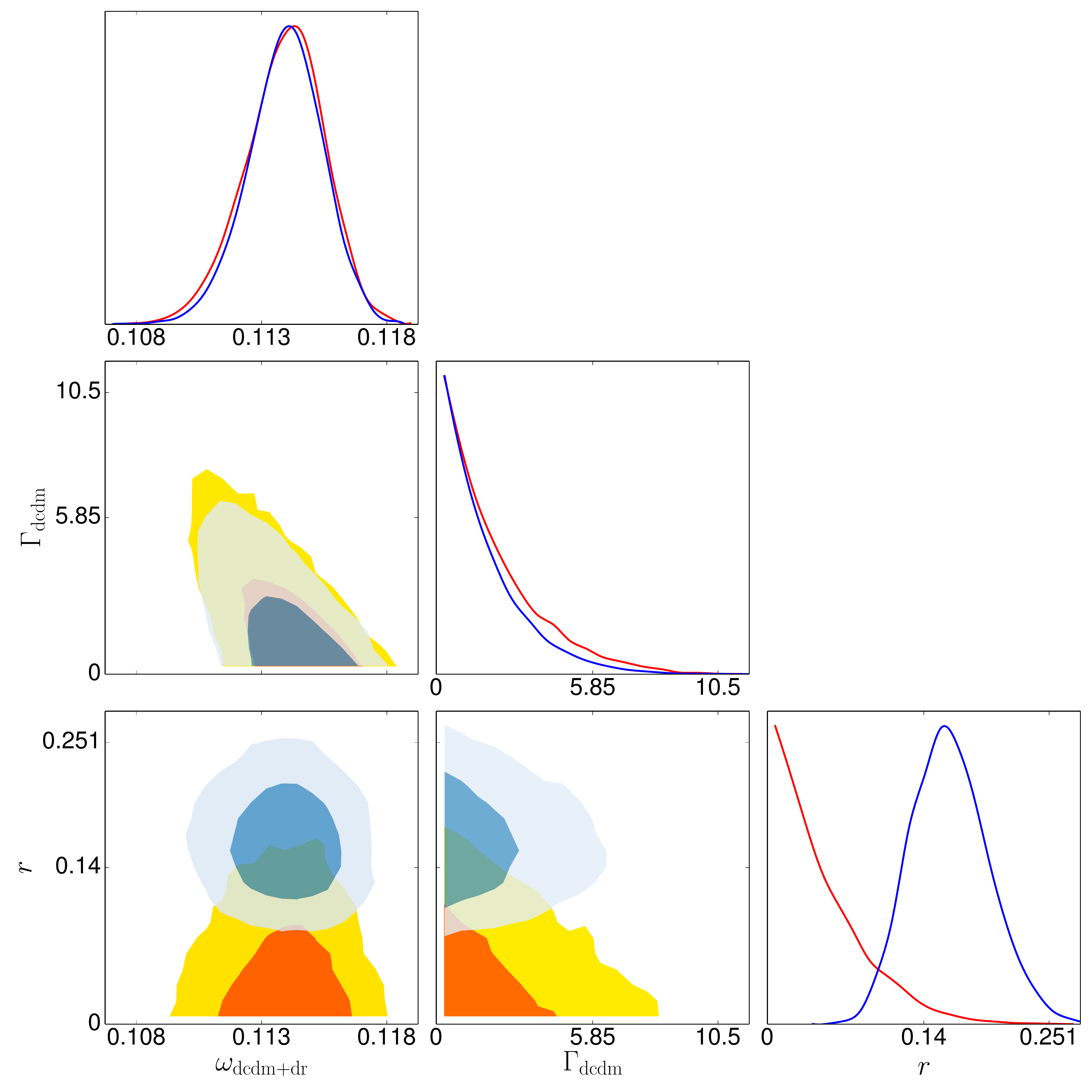}
\caption{Comparison of the results for $\{ \omega_\mathrm{dcdm+dr}, \Gamma_\mathrm{dcdm}, r \}$ for the $\Lambda$CDM + $\{\Gamma_\mathrm{dcdm}, r \}$ model for the 1-d and 2-d posterior distributions, using the dataset set $A$ (blue contours) and $B$ (yellow/orange contours). The contours represent 68\% and 95\% confidence levels.}\label{fig:triangle1}
\end{figure}

\begin{figure}[ht]
\centering
\includegraphics[width=0.9\textwidth]{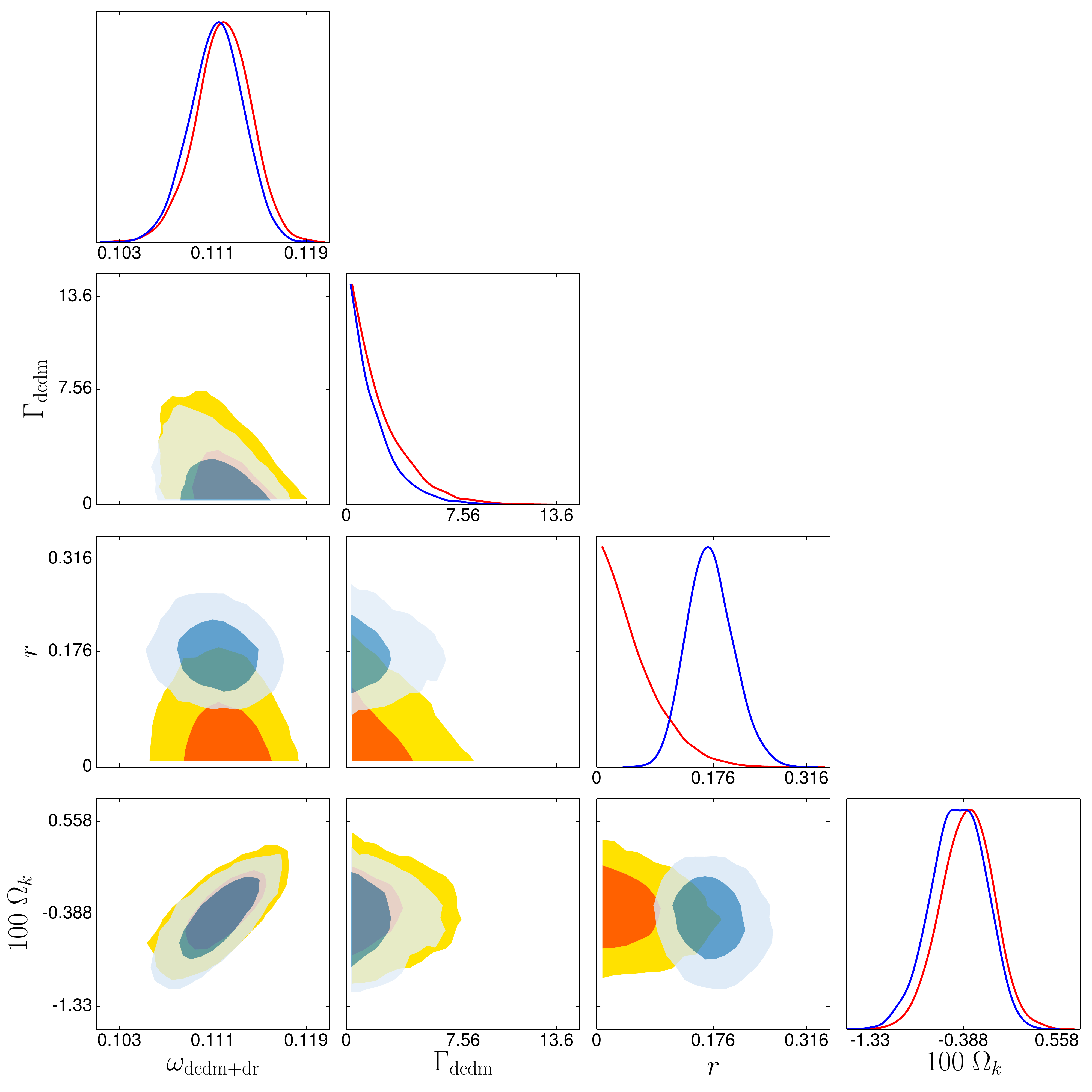}
\caption{For the $\Lambda$CDM + $\{\Gamma_\mathrm{dcdm}, r, \Omega_k \}$ model, comparison of the results for $\{ \omega_\mathrm{dcdm+dr}, \Gamma_\mathrm{dcdm}, r, \Omega_k \}$ using the dataset set $A$ (blue contours) and $B$ (yellow/orange contours), for the 1d and 2d posterior distributions. The contours represent 68\% and 95\% confidence levels.}\label{fig:triangle2}
\end{figure}

The results are summarized in Table~\ref{tab:results} and Figures~\ref{fig:triangle1} and  \ref{fig:triangle2}. 
\begin{table}[ht]
  \begin{center}
  \begin{tabular}{|c|cc|cc|}
  \hline 
  &&&& \\[-10pt]
  Model & \multicolumn{2}{c|}{$\Lambda$CDM + $\{\Gamma_\mathrm{dcdm}, r \}$} & \multicolumn{2}{c|}{$\Lambda$CDM + $\{\Gamma_\mathrm{dcdm}, r, \Omega_k\}$} \\[5pt]
  Data   & A & B & A & B\\[5pt]
  \hline \hline
  &&&& \\[-10pt]
  $100 \, \omega_\mathrm{b}$     & $2.231_{-0.024}^{+0.025}$    & $2.226_{-0.024}^{+0.024}$    & $2.247_{-0.030}^{+0.028}$    & $2.247_{-0.029}^{+0.028}$ \\[7pt]
  $H_0$ [km/s/Mpc]               & $68.89_{-0.61}^{+0.62}$      & $68.92_{-0.62}^{+0.61}$      & $68.21_{-0.79}^{+0.79}$      & $68.07_{-0.80}^{+0.83}$  \\[7pt]
  $10^9 A_s$                     & $2.145_{-0.050}^{+0.044}$    & $2.143_{-0.047}^{+0.044}$    & $2.157_{-0.054}^{+0.046}$    & $2.156_{-0.052}^{+0.045}$  \\[7pt]
  $n_s$                          & $0.9643_{-0.0056}^{+0.0055}$ & $0.9666_{-0.0056}^{+0.0055}$ & $0.9705_{-0.0077}^{+0.0071}$ & $0.9742_{-0.0076}^{+0.0072}$ \\[7pt]
  $\tau_\mathrm{reio}$           & $0.082_{-0.011}^{+0.012}$    & $0.082_{-0.011}^{+0.011}$    & $0.08676_{-0.013}^{+0.012}$  & $0.08792_{-0.013}^{+0.011}$ \\[7pt]
  $\omega_\mathrm{dcdm+dr}$      & $0.1142_{-0.0014}^{+0.0016}$ & $0.1142_{-0.0014}^{+0.0017}$ & $0.1117_{-0.0023}^{+0.0026}$ & $0.1113_{-0.0023}^{+0.0025}$ \\[7pt]
  $\Gamma_{\rm dcdm}$ [$\mbox{km s}^{-1} \mbox{Mpc}^{-1}$] & $< 5.9$    & $<5.0$               & $<6.0$                       & $<4.9$  \\[7pt]
  $r$                            & $<0.13$                      & $0.164_{-0.040}^{+0.032}$    & $0.05273_{-0.053}^{+0.012}$  & $0.1713_{-0.039}^{+0.033}$ \\[7pt]
  $10^2 \Omega_k$                & --                           & --                           & $-0.3517_{-0.26}^{+0.28}$    & $-0.4405_{-0.27}^{+0.30}$  \\[5pt]
  \hline 
  &&&& \\[-10pt]
  $\tau_{\rm dcdm}$ [Gyr]        & $>160$                       & $>200$                       & $>160$                       & $>200$ \\[5pt]
  \hline 
  \end{tabular}
  \end{center}
  %
  \caption{Marginalised Bayesian credible intervals for the cosmological parameters of the models considered in our analysis. We quote either mean values and 68\% confidence levels or 95\% upper/lower bounds. 
  The last lines show the results for the derived parameter $\tau_{\rm dcdm} = 1/\Gamma_{\rm dcdm}$ representing the dcdm lifetime (assuming a flat prior on the rate $\Gamma_{\rm dcdm}$, and not on the lifetime).}
  \label{tab:results}
\end{table}

For the $\Lambda$CDM + $\{\Gamma_\mathrm{dcdm}, r \}$ model, we find that the best-fit model has a negligible decay rate. Using the $A$ dataset, the upper bound is $\Gamma_{\rm dcdm} < 5.9\,\mbox{km s}^{-1} \mbox{Mpc}^{-1}$ (95\% CL). The decay rate is not significantly correlated with any other cosmological parameter, except $\omega_\mathrm{dcdm+dr}$ and $r$, as can be seen in Figure~\ref{fig:triangle1}. Indeed, the data prefer a certain amount of DM at early times, corresponding to the correct redshift of equality. Hence models with a large decay rate have a smaller DM density today, explaining the negative correlation between  $\Gamma_{\rm dcdm}$ and $\omega_\mathrm{dcdm+dr}$. There is also a correlation between $\Gamma_{\rm dcdm}$ and $r$: both parameters can enhance the small-$l$ CMB temperature spectrum, so larger values of $r$ lead to a stronger bound on $\Gamma_{\rm dcdm}$. Still, since $r$ is peaked in zero (as usual using Planck data), we know that the bound on $\Gamma_{\rm dcdm}$ that we would obtain under the assumption $r=0$ would be very similar to what we get here. 

For the same model and the $B$ dataset, the bounds on the tensor-to-scalar ratio moves close to $r\simeq 0.17$ at $k_*=0.05$/Mpc (slightly lower than in the $\Lambda$CDM + $r$ model, because of the correlation with $\Gamma_{\rm dcdm}$), pushing the bound on the DDM decay rate down to $4.8\,\mbox{km s}^{-1} \mbox{Mpc}^{-1}$.

For the $\Lambda$CDM + $\{\Gamma_\mathrm{dcdm}, r, \Omega_k\}$ model, and using either the $A$ or $B$ data set, we see in Figure~\ref{fig:triangle2} that $\Gamma_{\rm dcdm}$ is not correlated with $\Omega_k$, and that the bounds on $\Gamma_{\rm dcdm}$ are nearly the same as in the flat model. Indeed, the combination of CMB and LSS data allow us to distinguish between the effects of these two parameters. Since $r$ and $\Omega_k$ are the two parameters most likely to be degenerate with $\Gamma_{\rm dcdm}$ within the simplest extensions of $\Lambda$CDM, we conclude that current CMB and LSS data provide very robust limits on the dcdm decay rate, depending on the data set, but not on the assumed cosmological model.

\section{Conclusions}\label{sec:conclusions}
We have shown that the lifetime of CDM must be above $160\,\text{Gyr}$ (or $200\,\text{Gyr}$ assuming that BICEP2 has detected gravitational waves), even for the most conservative case where CDM decays entirely into Dark Radiation. This is a model independent bound, since it relies only on the gravitational interactions of CDM and its decay product which can not be avoided in any particle physics model. If the decay product is allowed to have a mass, we would expect this bound to worsen depending on the mass of the daughter particles. We will consider this scenario in a future publication.

The bound on $\Gamma_{\rm dcdm}$ has relevant implications on particle physics model buildings. Depending on the specific scenario containing a decaying massive particle which may act as a DM contribution, the lifetime constraint can typically be translated into a lower bound on the particular new mass scale which enters the decay process via a non standard interaction. As a key example, consider the already mentioned Majoron scenario. In this case the pseudo--scalar Goldstone boson related to the breaking of lepton number conservation, acquires a finite mass due to non-perturbative quantum gravity effects that explicitly break global symmetries, and decays into (mainly) neutrino pairs. From its decay rate of Eq. (\ref{majorondecayrate}), a lifetime larger than 200 Gyr translates into the following lower bound on the lepton number breaking scale $v$ 
\begin{equation}
 v > 4.4  \cdot 10^8\, \frac{m_\nu}{\mbox{ eV}}  \left(\frac{m_J}{\mbox{keV}} \right)^{1/2}\,\,\mbox{GeV}\,.
\end{equation}
This is just an example of how a strong constraint on DM stability can provide relevant information on its yet unknown nature and constrain models of new non-standard interactions. Finally, we would like to remark that these bounds are expected to become even stronger in the near future. Indeed, a key role in their improvement will be  played by future weak lensing surveys, which will also help in reducing  degeneracies with massive neutrinos, see e.g.~\cite{Wang:2010ma,Wang:2012eka}.

\section*{Acknowledgements}
GM acknowledges support by the {\it Istituto Nazionale di Fisica Nucleare} I.S. TASP and by MIUR, PRIN {\it Fisica Teorica Astroparticellare}.
BA, JL and TT received support from the Swiss National Foundation. At LAPTh, this activity was developed coherently with the
research axes supported by the ANR Labex grant ENIGMASS.
\bibliographystyle{utcaps}
\bibliography{decayDM}


\end{document}